\documentstyle[12pt,epsfig,rotating]{article} 
\makeatletter
\def\section{\@startsection {section}{1}{\z@}{-1cm plus-1ex
    minus-.2ex}{1.5ex plus.2ex}{\reset@font\normalsize\bf}}
\def\subsection{\@startsection{subsection}{2}{\z@}{-0.7cm plus-1ex
    minus-.2ex}{1.0ex plus.2ex}{\reset@font\normalsize\it}}
\def\subsubsection{\@startsection{subsubsection}{3}{\z@}{-0.7cm plus-1ex 
    minus-.2ex}{1.0ex plus.2ex}{\reset@font\normalsize}}
\def\paragraph{\@startsection
     {paragraph}{4}{\z@}{3.25ex plus1ex minus.2ex}{-1em}{\reset@font
     \normalsize\bf}}
\def\subparagraph{\@startsection
     {subparagraph}{4}{\parindent}{3.25ex plus1ex minus
     .2ex}{-1em}{\reset@font\normalsize\bf}}

\ifcase \@ptsize
\or 
\or 
\fi
\advance\textheight by \topskip
\advance\footskip by 6mm

\newcount\dfttsuboption 
\def\dfttnum#1{\def\@dfttnum{#1}}
\dfttnum{??/xx}
\def\dfttmemo#1{\dfttsuboption=2\def\draftname{INTERNAL MEMO #1\ }}

{ \newcount\hour \newcount\minutes \newcount\hourint%
  \minutes=\time \hour=\time \divide\hour by 60%
  \hourint=\hour   \multiply\hourint by -60  \advance\minutes by \hourint%
  \xdef\@time{\the\hour:\ifnum\minutes>9\else0\fi\the\minutes}%
}

\def\@maketitle{\newpage
 \null\begingroup\def\baselinestretch{1}
\raggedleft\normalsize DFTT 50/96 \\ 
\raggedleft\normalsize MPI-PhT/96-80 \\
\raggedleft\normalsize LU TP 96-23  \\
\raggedleft\normalsize August 28th, 1996 \\
 \vskip 0.8cm
 \begin{center}%
  {\Large \@title \par}%
  \vskip 1.5em
  {\normalsize
   \lineskip .5em
   \begin{tabular}[t]{c}\@author
   \end{tabular}\par}%
  \ifnum\dfttsuboption=0 \vskip 1em{\footnotesize \@date}\fi%
 \end{center}%
 \par\endgroup
 \vskip 1cm}

\newenvironment{summary}{\begin{quote}\begin{center}\bf\abstractname\par%
\end{center}\vskip 0.25em}{\end{quote}\vskip 2em}

\def\eref#1{(\ref{#1})}
\makeatother

\dfttmemo{1}

\newcommand{\be}{\begin{equation}}
\newcommand{\ee}{\end{equation}}
\newcommand{\ba}{\begin{eqnarray}}
\newcommand{\ea}{\end{eqnarray}}

\newcommand\pp{$\pm$}

\newcommand{\jetset}{{\sc Jetset}}
\newcommand{\delphi}{{\sc Delphi}}

\newcommand{\NF}{\cal{N}_{\kern -1.9pt f}}     
\newcommand{\NC}{\cal{N}_{\kern -1.7pt c}}     

\newcommand{\pT}{{p\kern -.2pt\lower 4pt\mbox{\tiny T}}}    

\newcommand{\pL}{{p\kern -.2pt\lower 4pt\hbox{\tiny L}}}    

\newcommand{\ymin}{y_{\mathrm{min}}}         

\title{ \bf The negative binomial 
 distribution in quark jets with fixed flavour  
  \thanks{\it Work supported in part by M.U.R.S.T. (Italy) under grant 1995.} }

\author{A.\  GIOVANNINI$^1$ \thanks{E-mail: giovannini@to.infn.it}\ , \ 
 S.\  LUPIA$^2$ \thanks{E-mail: lupia@mppmu.mpg.de}\ , \ 
 R.\ UGOCCIONI$^3$ \thanks{E-mail: roberto@thep.lu.se} \\ \\ 
\it $^1$ Dip. Fisica Teorica and I.N.F.N. -- Sezione di Torino, \\ 
\it via Giuria 1, I-10125 Torino, Italy  \\  \\  
\it $^2$  Max-Planck-Institut f\"ur Physik (Werner-Heisenberg-Institut), \\
\it F\"ohringer Ring 6, D-80805 M\"unchen, Germany \\  \\  
\it $^3$ Dept. of Theoretical Physics, University of Lund, \\
\it S\"olvegatan 14 A, S 223 62, Lund, Sweden}

\begin{document}

\maketitle

\vspace{-1.0cm} 

\begin{summary} 
We show that 
both the multiplicity distribution and the ratio of factorial cumulants 
over factorial moments for 2-jet events 
in $e^+e^-$ annihilation at the $Z^0$ peak can 
be well reproduced by the weighted superposition of two negative
binomial distributions, associated to the contribution  of $b\bar b$ 
and light flavoured events respectively. 
The negative binomial distribution is then suggested to describe the  
multiplicity distribution of 2-jet events with fixed flavour.
\end{summary}

\vspace{-0.5cm} 
PACS: 13.65

\newpage

\section{Introduction}

Two different experimental effects in the 
Multiplicity Distributions (MD's)  of charged particles  
in full phase space in  $e^+e^-$ annihilation at the $Z^0$ peak, 
i.e., 
the shoulder  visible in the intermediate multiplicity 
range\cite{delphi:2,opal,aleph} and the quasi-oscillatory behaviour of 
the ratio of factorial cumulants over factorial 
moments of the MD, $H_q$, when plotted as a function of its order $q$
\cite{sld,Gianini}, have been quantitatively reproduced in \cite{hq2} 
 in terms of a weighted superposition of two Negative Binomial Distributions 
 (NBD's), associated  to two- and multi-jet events respectively.
A further test of this picture, in which 
the simple NBD appears at a very elementary level of
investigation, is provided by the study of 
samples of events with a fixed number of jets.
In \cite{delphi:3}, the \delphi\ Collaboration has shown that a single NBD can
describe the MD's for events with a fixed number of jets for a range of
values of the jet resolution parameter $\ymin$. 
This was indeed the starting point of the parametrization proposed in
\cite{hq2}. 

In this letter, by extracting the ratio $H_q$  from
published MD's according to the procedure described in \cite{hq2}, we show 
that the oscillations observed experimentally are larger than those
predicted by a single NBD, even after taking into account the truncation effect,
which was shown\cite{hq} to be important in the behavior of $H_q$'s. 
These results suggest that, while
hard gluon radiation plays a relevant role in the
explanation of the shoulder structure of MD's and of oscillations of the ratio
$H_q$, some other effects should be taken into account 
for a detailed description of 
experimental data of events with a fixed number of jets. 
In this respect, it is worth recalling the interesting results obtained by the
OPAL Collaboration\cite{opalfb} on forward-backward correlations and on 
the increase of transverse momentum of produced hadrons in the intermediate
multiplicity range: it has been found indeed that both effects are mainly due
to hard gluon radiation, i.e., to the superposition of events with a fixed
number of jets. However, a residual positive correlation has been found in
a sample of 2-jet events; via Monte Carlo simulations, this effect has been
associated to the combined action of superposition of events with different
quark flavours and, in the central region, to a residual effect due to
resonances' decays. 
Let us remind, however, that 
the presence of heavy flavours has been shown not to 
affect the increase of the transverse momentum of produced hadrons in the
intermediate multiplicity range, thus suggesting that not all observables are
sensitive to the original quark flavour. 
A theoretical study based on Monte Carlo simulations has first 
suggested that 
the study of MD's can indeed point out interesting features of particle
production in $b\bar b$ events\cite{GUV}.
Recently the \delphi\ Collaboration has established experimentally 
the sensitivity of MD's to the original quark flavour, by comparing the 
MD for the full sample of events with
a sample enriched in $b\bar b$ events\cite{delphibb}.

We propose 
in this letter  to associate a NBD to the MD in 2-jet events of fixed flavour.
 We show, after examining possible alternatives,  
 that the  weighted superposition of two NBD's,  
which we associate to $b\bar b$ and light flavoured events, 
describes very well both the MD's and the ratio $H_q$ for 2-jet events; the two
NBD's have the same  $k$ parameters and differ 
in the average multiplicity only. 
Some consequences of this fact are examined in the conclusions.

\section{MD's in $b$-jets}

The \delphi\ Collaboration has studied the effect of quark
flavour composition on the MD in one hemisphere, by comparing the 
MD for the full sample of events with
that for a sample enriched in $b\bar b$ events\cite{delphibb}: 
the MD extracted from the $b\bar b$ sample 
was found to be essentially identical in shape to the MD obtained for 
the full sample, apart from a shift of one unit, which may be related to the
effect of weak decays of $B$-hadrons\cite{dias}. 
To give a quantitative comparison of the MD's in 
single hemisphere in  the $b\bar b$ sample 
and in the sample with a mixture of all flavours, 
we have fitted both experimental MD's   with a  NBD and with a NBD shifted by
one or two units.  
The results of the fit are shown in Table~\ref{singleb}. 
A single NBD gives a poor description of the MD for the $b\bar b$ sample; 
the description improves strongly if one introduces a shift by one unit, and
becomes even better with a shift by two units. 
The reason is that with the shift the NBD is able to reproduce better
the head of the distribution; however the tail remains underestimated. 
In this respect, one should 
remember that the single NBD cannot give a good description 
 of the MD with all flavours in full phase space, 
since it cannot reproduce the shoulder structure
 due to the superposition of events with different number of
 jets\cite{delphi:3}. The fact that this feature should be present for 
the $b\bar b$ sample too should be verified experimentally.

\begin{table}    
\caption[table:singleb]{Parameters and $\chi^2$/NDF of the 
fit to experimental data\cite{delphibb} on single
hemisphere MD's for $b\bar b$ sample and for all flavours 
with a single NBD and with a NBD shifted by one or two
 units. }\label{singleb}
 \begin{center}
 \vspace{4mm}
 \begin{tabular}{||c|c|c||}
\hline
  & $b\bar b$ sample & all flavours \\ \hline
\multicolumn{3}{||l||}{NBD}  \\ \hline 
$\bar n$ & 11.67\pp 0.07 & 10.67\pp 0.02 \\ 
$k$ &  24\pp 2 & 14.5\pp 0.3  \\
$\chi^2$/NDF & 118/26 & 140/28  \\  \hline
\multicolumn{3}{||l||}{NBD (shift by 1 unit)}  \\ \hline 
$\bar n$ & 10.62\pp 0.07 &  9.63\pp 0.02 \\ 
$k$ &  15.8\pp 0.6 & 10.42\pp 0.2  \\
$\chi^2$/NDF & 60/26 & 64/28  \\  \hline
\multicolumn{3}{||l||}{NBD (shift by 2 units)}  \\ \hline 
$\bar n$ & 9.55\pp 0.07 &  \\ 
$k$ &  10.4\pp 0.3 &   \\
$\chi^2$/NDF & 20/26 &   \\  \hline
 \end{tabular}
 \end{center}
\end{table} 

In any case, interesting information on the properties of these MD's 
can be extracted  without using any parametrization at all. In what follows
we will consider only 2-jet events, selected with a suitable algorithm,
but the same reasoning can be carried out also for the full sample.
Let us call $p_n$,  $p^b_n$ and $p^l_n$ 
the MD's in a single hemisphere for all events,
for $b\bar b$ events and for light flavoured (non $b \bar b$) events
respectively, and $g(z) \equiv \sum_{n=0}^{\infty} p_n z^n$, $g^b(z)$ and
$g^l(z)$ the associated generating functions. 
With $\alpha$ the fraction of $b\bar b$ events, one  has:
\be
  p_n = \alpha p^b_n + (1-\alpha) p^l_n					
\ee
i.e., 
\be
  g(z) = \alpha g^b(z) + (1-\alpha) g^l(z)	
\label{peso}
\ee
These relations are valid in general; \delphi\ data and our analysis shown in
Table~\ref{singleb} suggest that  $p^b_n$ is given by
 $p_n$ with a shift of one unit: 
 \be
  p^b_n = p_{n-1} \qquad  n > 0; 
  \qquad\qquad         p^b_0 = 0;			\label{pshift}
\ee
i.e., 
\be
  g^b(z) = z g(z)          
  \label{shift}                                      
\ee
Substituting now eq.~\eref{shift} in eq.~\eref{peso}, one gets:
\be
  g(z) = \frac{1-\alpha}{1-\alpha z} g^l(z)      
  \ee
i.e.,
\be
  g^b(z) =  \Bigl[ \frac{z (1-\alpha)}{1-\alpha z} \Bigr]  g^l(z)                           
\label{result}
\ee
The MD in $b\bar b$ events 
is the convolution of  a shifted geometric distribution, 
of  average value $1/(1-\alpha)$, with the MD in light flavoured events. 
The shifted geometric MD could be related to the MD of the decay products 
of B hadrons in the framework of  \cite{dias}. 

The connection of the MD in a  single hemisphere to that 
in full phase space is not entirely trivial, since one has to take into
account additional effects, like for instance charge conservation,  
which requires that the final multiplicity be even. 

Let us use the same notation for MD's as in the previous paragraph, but with
capital letters to denote the MD's in full phase space; 
by taking the two hemispheres as independent (which, as suggested in
\cite{opalfb}, should be a good approximation at least for light flavours)
but applying charge conservation, we obtain:
\ba
  P^i(n_1,n_2) = \cases{ 2 p^i_{n_1} p^i_{n_2} 
                    &if $n_1+n_2$ is even  \cr
                 0  & \hbox{otherwise} \cr}                     
\ea
Here  $P^i(n_1,n_2)$ is  
 the probability to produce $n_1$ particles in one hemisphere and
$n_2$ in the other hemisphere and  $i$ denotes 
either all 2-jet events (no label), 
or $b\bar b$ events ($i=b$) or light flavoured
events ($i=l$). The factor 2 is for normalization, assuming that the
$p_n^i$ do not privilege the even or odd component; in any case this
effect can be easily taken into account and results similar to those
given below are obtained.

The MD in full phase space is given by definition by: 
\be                   
P_n^i  =  \sum_{n_1=0}^n  P^i(n_1, n-n_1)	
\ee 
In terms of the generating functions, one obtains
\ba
  G^i(z) &=& \sum_{n=0}^{\infty}   z^n  P^i_n = 
	2 \sum_{\scriptstyle n=0 \atop\scriptstyle n \mathrm{~even}}^{\infty} 
  \sum_{n_1=0}^n  z^{n_1}   p^i(n_1) z^{n-n_1}    p^i(n-n_1) \nonumber \\ 
        &=& \bigl( [g^i(z)]^2  + [g^i(-z)]^2  \bigr)				
\label{fps}
\ea

We can see now how the relations which we obtained in the
previous paragraph are modified going from a single hemisphere to full phase
space: by putting eq.~\eref{result} in eq.~\eref{fps}, one has 
\be
  G^b(z) = z^2 \frac{ (1-\alpha)^2}{(1-\alpha z)^2} [g^l(z)]^2 
  + z^2 \frac{ (1-\alpha)^2}{(1+\alpha z)^2} [g^l(-z)]^2 
\label{fps2}
\ee
For $\alpha$ = 0, one obtains that in full phase space 
the MD in $b\bar b$ events  coincides with the MD for light flavoured events
with a shift of two units, as we get $G^b(z) = z^2 G^l(z)$.
The MD for small values of $\alpha$ should not be too far from this limit, as
can be easily checked with numerical examples. 

In conclusion,
going from single hemispheres to full phase space by taking into account only
charge conservation, the MD of $b\bar b$ events becomes close not to the total 
MD but to the MD of light flavoured events. 
The two MD's seem to have the same
characteristics, like average value and dispersion, 
and the only difference should lie in
a shift of two units.

\section{MD's and $H_q$'s ratio in 2-jet events}

\begin{figure}
\begin{center}
\mbox{\begin{turn}{90}%
\epsfig{file=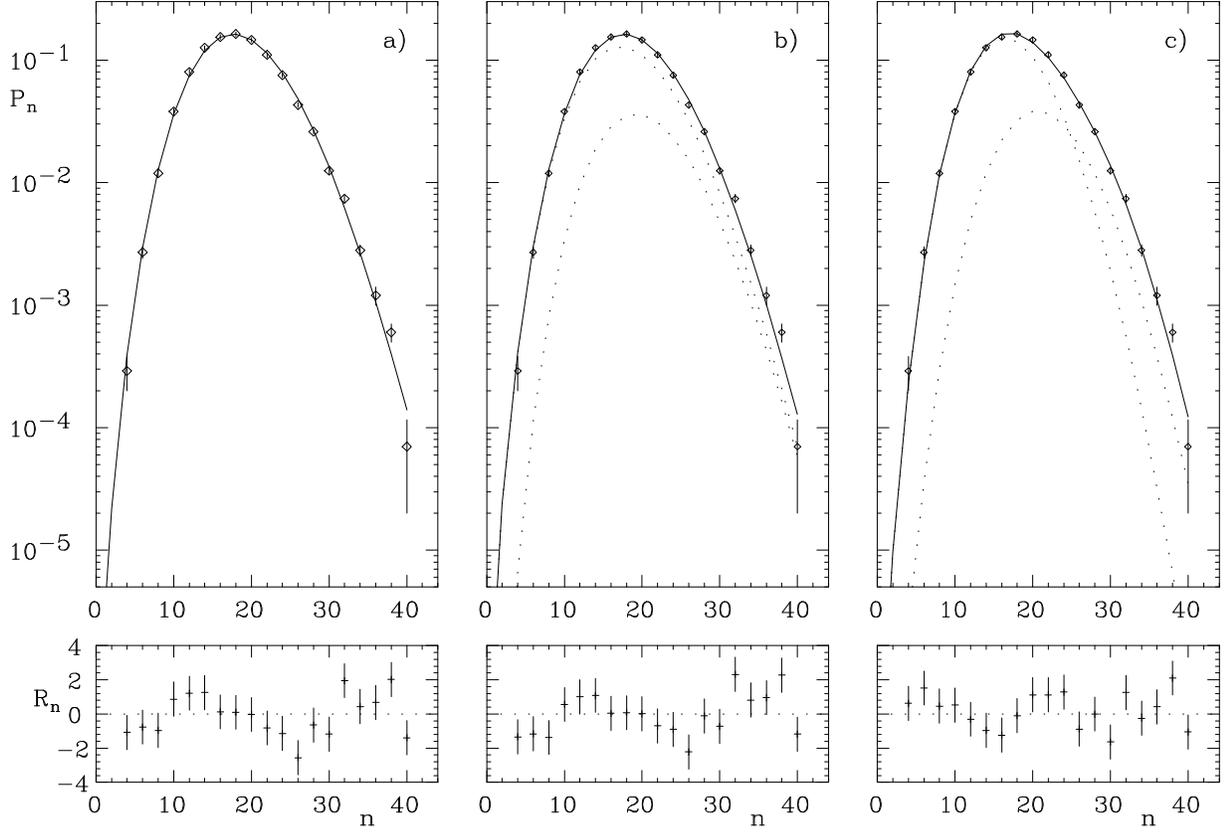,bbllx=0.cm,bblly=2.cm,bburx=22cm,bbury=26.cm,width=15cm}
\end{turn}}          
\end{center}
\caption[figure:fit]{%
{\bf a)}: charged particles' MD for two-jet events 
in full phase space, $P_n$, at the $Z^0$ peak  from 
\delphi\cite{delphi:3} with $\ymin$ = 0.02 are 
compared with the fit with a single NBD as 
performed by \delphi\ Collaboration  (solid lines); 
the lower part of the figure shows the residuals, $R_n$, i.e., 
the difference between data and theoretical predictions, 
expressed in units of standard deviations. 
The even-odd effect has been taken into account (see eq.~\protect\eref{pnfps}). 
{\bf b)}: 
Same as in {\bf a)}, but the solid line here shows
the result of fitting eq.~\protect\eref{2par},
with parameters given in Table~\protect\ref{fits}. 
{\bf c)}: 
Same as in {\bf a)}, but the solid line here shows
the result of fitting eq.~\protect\eref{poisson},
with parameters given in Table~\protect\ref{fits}.} 
\label{fit}
\end{figure}

The analysis of MD's for events with a fixed number of jets, and in particular 
2-jet events,  has been performed in \cite{delphi:3}, 
where a single NBD has been shown to 
reasonably describe  the MD's for events with a fixed number of jets, for
several values of the jet resolution parameter $\ymin$
(the JADE jet-finding algorithm has been used in  \cite{delphi:3}).
A comparison of \delphi\ data with $\ymin$ = 0.02 with a single NBD
is shown in Figure~\ref{fit}a together with the residuals, i.e., the
normalized difference between data and theoretical predictions, which 
 point out the presence of substructures in experimental data. 
In view of previous results, it is then interesting to investigate whether 
these substructures can be explained in terms of 
the different contribution of quark-jets with different flavours. 

\begin{table}    
 \caption[table:fits]{Parameters and $\chi^2$/NDF of 
the fit to experimental data on 2-jet events
MD's from \delphi\cite{delphi:3} with three different MD's:
the weighted superposition of a NBD and a 
shifted NBD with the same parameters (eq.~\eref{2par}), 
the weighted superposition of a Poisson plus a 
shifted Poisson (eq.~\eref{poisson})
and the weighted superposition  of two NBD's with the 
same $k$ (eq.~\eref{3par}). 
The weight used is the fraction of $b\bar b$ events. 
The even-odd effect has been taken into account (see eq.~\protect\eref{pnfps}). 
Results are shown for different values of  the jet-finder parameter $\ymin$.} 
\label{fits}
 \begin{center}
 \begin{tabular}{||c|c|c|c||}
\hline
   &  $y_{min}$ = 0.01  & $y_{min}$ = 0.02  & $y_{min}$ = 0.04 \\ \hline
\multicolumn{4}{||l||}{NBD + shifted NBD (same $\bar n$ and $k$) 
eq.~\eref{2par}}  \\ \hline 
$\bar n$ & 17.17\pp 0.05 & 18.01\pp 0.04 & 18.99\pp 0.05 \\ 
$k$ & 69\pp 5 & 57\pp 3 & 44\pp 1 \\
$\chi^2$/NDF & 18.2/17 & 27.9/17 &  53.9/21  \\  \hline
\multicolumn{4}{||l||}{Poisson + shifted Poisson eq.~\eref{poisson}}  \\ \hline 
$\bar n_1$ & 19.67\pp 0.13 & 21.10\pp 0.10 & 22.83\pp 0.09 \\ 
$\bar n_2$ & 16.37\pp 0.07 & 16.95\pp 0.06 & 17.51\pp 0.06 \\
$\chi^2$/NDF & 21.7/17 & 20.75/17 &  45.05/21  \\  \hline
\multicolumn{4}{||l||}{2 NBD (same $k$) eq.~\eref{3par}}  \\ \hline 
$\bar n_l$ & 16.81\pp 0.21 & 17.22\pp 0.15 & 17.98\pp 0.15 \\ 
$\bar n_b$ & 20.26\pp 1.71 & 21.96\pp 1.57 & 23.61\pp 1.64 \\ 
$k$ & 124\pp 51 & 145\pp 53 & 120\pp 33 \\
$\chi^2$/NDF & 17.4/16 & 12.6/16 &  27.5/20  \\
$\delta_{bl}$ & 3.44\pp 0.83 & 4.6\pp 0.5 & 5.6\pp 0.5 \\ \hline
 \end{tabular}
 \end{center}
\end{table} 

We  parametrize the
experimental data on MD's for 2-jet events 
in full phase space in $e^+e^-$ 
annihilation at the $Z^0$ peak\cite{delphi:3} 
in terms of the superposition of 2 NBD's, 
associated to the contribution of 
$b$- and light flavours (we include the charm among the light flavours 
 to a first extent). 
We fix therefore the weight parameter 
to be equal to the fraction of  $b \bar b$ events,  
 $\alpha$ = 0.22\cite{rb}. 
Following the results of the previous section, 
we ask that the NBD associated to the $b$ flavour be shifted by two units and
that both parameters of the two NBD's, $\bar n$ and $k$, be the same. 
Formally, we perform then a fit  with the following 2-parameter distribution: 
\be
P_n(\bar n, k) = \alpha 
P_{n-2}^{\mathrm{NB}}(\bar n, k) + (1 - \alpha ) P_n^{\mathrm{NB}}(\bar n, k)
\label{2par} 
\ee
where $P_{n-2}^{\mathrm{NB}}(\bar n, k) = 0$ for $n < 2$.
$P_n^{\mathrm{NB}}(\bar n,k)$  is here the NBD, 
expressed in terms of two parameters, the average multiplicity $\bar n$ 
and the parameter $k$, linked to the dispersion by 
$D^2/\bar n^2 = 1/\bar n + 1/k$,  as: 
\be
P_n^{\mathrm{NB}}(\bar n, k) = \frac{k(k+1)\dots (k+n-1)}{n!} 
\left( \frac{k}{\bar n +k} \right)^k  
 \left( \frac{\bar n}{\bar n + k} \right)^n
\ee
As far as MD's in full phase space are concerned, one has also 
to take care of the ``even-odd'' effect, 
i.e., of the fact that the total number of  final charged particles 
must be even due to charge conservation; 
accordingly, the actual form used in the fit procedure is given by: 
\be
   P_n^{fps} = \cases{ A P_n 
                    &if   $n$ is even  \cr
                 0  & \hbox{otherwise} \cr}                     
\label{pnfps}
\ee
where $A$ is the normalization
parameter, so that $\sum_{n=0}^{\infty} P_n^{fps} = 1$. 

Table~\ref{fits} (first group)
shows the two parameters $\bar n$ and $k$ of eq.~\eref{2par}
for different values of the jet resolution parameter $\ymin$;
the proposed parametrization gives a rather good description with only two
parameters; the agreement is worse for $\ymin$ = 0.04,
which could be due to the contamination of 3-jet events. 
However, as shown in
Figure~\ref{fit}b for $\ymin$ = 0.02,  
the oscillatory structure in the residuals does not disappear
with this parametrization.
Let us remind that the values of $\chi^2$/NDF shown in the Table 
should be considered just indicative, 
since we did not  know the full covariance matrix and we 
could not then treat properly the
correlations between different channels of the MD. 
This forbids also a direct comparison of the $\chi^2$/NDF of the present
parametrization with the values of $\chi^2$/NDF for a single NBD 
obtained in \cite{delphi:3}, where correlations between bins
were taken into account.
Finally, let us say that we also fit eq.~\eref{fps} by assuming that
$p^l_n$ is a NBD, and we found the same results.

It is interesting at this point to investigate
a minimal model where no physical correlations are  present
at the level of events with fixed number of jets and fixed flavour: we have
performed a fit using a weighted sum of a shifted 
Poisson plus a Poisson distribution
(with the correction for the even-odd effect according to eq.~\eref{pnfps}): 
\be
P_n(\bar n_1, \bar n_2) = \alpha 
P_{n-2}^{\mathrm{P}}(\bar n_1) + (1 - \alpha ) P_n^{\mathrm{P}}(\bar n_2)
\label{poisson} 
\ee
where 
\be
P_n^{\mathrm{P}}(\bar n) = \frac{\bar n^n}{n!} e^{-\bar n} 
\ee
Also in this case, we have two free parameters. 
Results of the fit are shown in Table~\ref{fits} (second group); 
also in this case a
reasonable fit is achieved, even though the MD at $\ymin=0.04$ 
shows again some anomaly. 
The MD with the parameters shown in Table~\ref{fits} 
is compared to experimental data with $\ymin$ = 0.02 in 
Figure~\ref{fit}c; it should be pointed out that the 
structure in the residuals is present in this case too. From 
this analysis, one would then conclude that physical 
correlations visible in
$e^+e^-$ annihilation result trivially from the superposition of samples of 
events with different quark flavours. 
A more accurate analysis with full covariance matrix
is needed to see which parametrization is preferred by experimental data. 
However, independent of the chosen parametrization, 
two different components, 
which can be associated to $b$- and light flavours contributions, 
are visible in the MD of 2-jet events. 

A more detailed analysis of the tail of the MD, which can help distinguish
different parametrizations,  comes from the
study of the ratio of unnnormalized factorial cumulant over
factorial moments 
\be
  H_q = \frac{\tilde K_q}{\tilde F_q}                                  \label{hmom}
\ee
as a function of the order $q$. 
The factorial moments, $\tilde F_q$, and
factorial cumulant moments, $\tilde K_q$,  can be obtained from the MD,
$P_n$, through the relations: 
\be 
 \tilde F_q = \sum_{n=q}^{\infty} n(n-1)\dots(n-q+1) P_n  ,         \label{facmom}
\ee
and 
\be
 \tilde K_q = \tilde F_q - 
 \sum_{i=1}^{q-1} {q-1 \choose i} \tilde K_{q-i} \tilde F_i .  
 \label{faccum}
\ee

 Since the $H_q$'s were shown to be sensitive to the truncation of the
tail due to the finite statistics of data samples\cite{hq}, moments have to be  
actually extracted from a truncated MD defined as follows 
(including again the correction for the even-odd effect as 
in eq.~\eref{pnfps}): 
\be
   P_n^{trunc} = \cases{ A' P_n
                 &if ($n_{min} \le n \le n_{max}$) 
                 and \ $n$ is even  \cr
                 0  & \hbox{otherwise} \cr}                     
\label{pntrunc}
\ee
Here $n_{min}$ and $n_{max}$ are the minimum and  the maximum observed 
multiplicity and $A'$ is a new normalization parameter. 

\begin{figure}
\begin{center}
\mbox{\epsfig{file=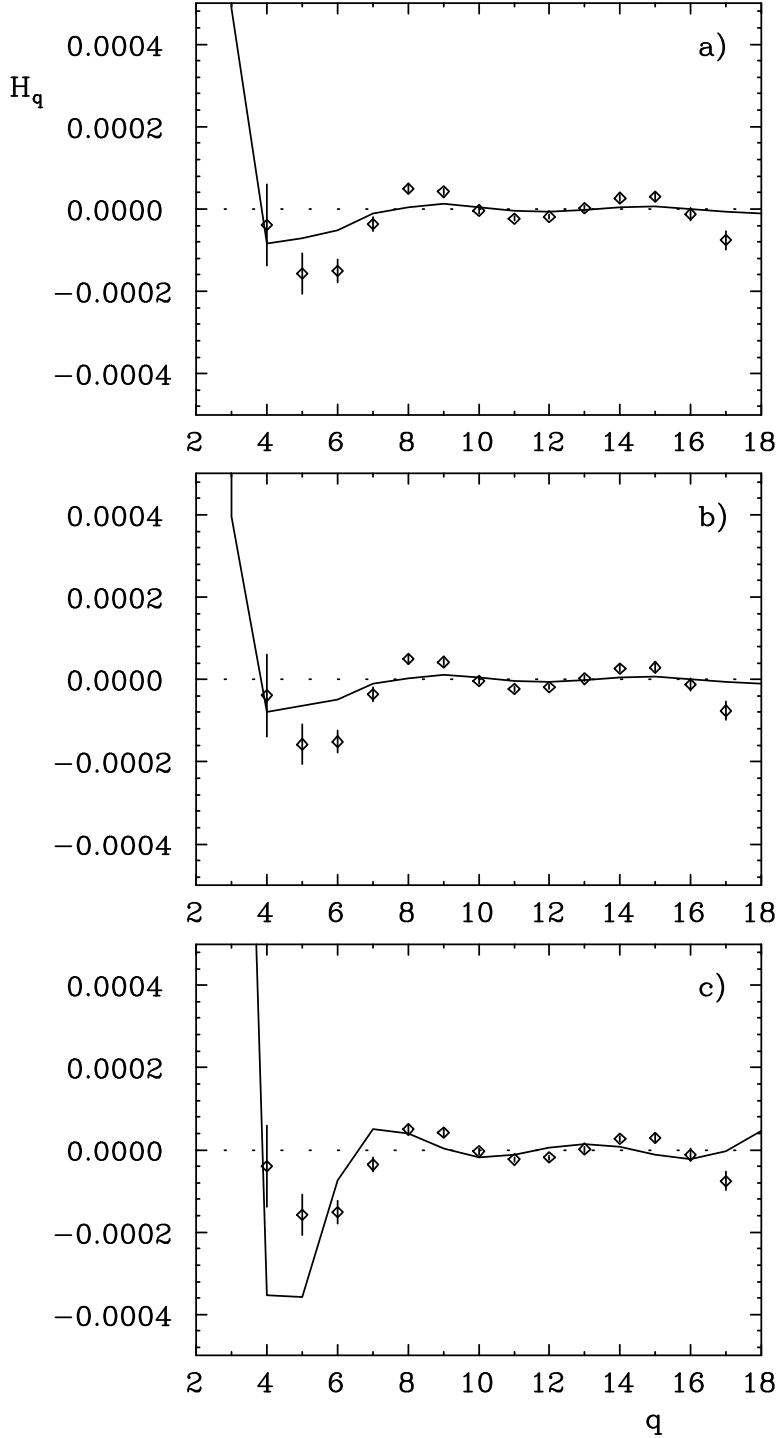,bbllx=4.cm,bblly=3.cm,bburx=16cm,bbury=27.cm,height=20cm}}
\end{center} 
\vspace{-0.5cm}
\caption[figure:hqfit]{The ratio of factorial cumulant over factorial moments, 
$H_q$, as a function of $q$. 
{\bf a)}: Experimental data (diamonds) for 2-jet events with $\ymin$ = 0.02 
are compared with the fit with a single NBD as 
performed by \delphi\ Collaboration (solid lines). 
{\bf b)}: 
Same as in {\bf a)}, but the solid line here shows
the result of fitting eq.~\protect\eref{2par},
with parameters given in Table~\protect\ref{fits}. 
{\bf c)}: 
Same as in {\bf a)}, but the solid line here shows
the result of fitting eq.~\protect\eref{poisson},
with parameters given in Table~\protect\ref{fits}. 
In all three cases the even-odd and the truncation effects have 
been taken into account (see eq.~\protect\eref{pntrunc}).} 
\label{hqfit}
\end{figure}

In Figure~\ref{hqfit} the $H_q$'s extracted from the 
experimental MD published by \delphi\ Collaboration~\cite{delphi:3} 
(here $\ymin$ = 0.02) with the procedure explained in~\cite{hq2} 
are compared with the predictions of  a single NBD as fitted  by 
\delphi\ Collaboration~\cite{delphi:3}, and 
of eq.s~\eref{2par} and \eref{poisson}. 
It is clear that all three parametrizations fail to describe the experimental
behaviour of the ratios $H_q$, i.e., the description of the tail of the MD is 
not accurate. 
We then conclude that  a single NBD cannot describe 
accurately the MD in 2-jet events, as already suggested by the study of
residuals of MD's. The  superposition of two NBD's
with the same parameters turns out also to be inadequate; one concludes 
that the imposed constraints are too strong and that 
some additional differences between $b\bar b$ and light flavoured
events should be allowed. 
Finally, the observed deviation from the Poisson-like fit suggests
that there are indeed dynamical correlations beyond the purely 
statistical ones.

\section{A new parametrization of MD's in 2-jet events} 

In the previous 
paragraph, the difference between the average multiplicity 
in $b\bar b$ and light flavoured events in 
the parametrization~\eref{2par}  has been  fixed 
to 2; by using the parametrization~\eref{poisson}, where this difference is not a priori
constrained, larger values have been obtained. Let us also remind that  
the experimental value of this observable at the $Z^0$ peak 
 is close to 2.8\cite{opalmult}; 
theoretical predictions in the framework of Modified Leading Log Approximation 
plus Local Parton Hadron Duality\cite{MLLAmult} give even larger values. 
It is therefore interesting to investigate whether one can reproduce not only 
the shape of the MD, but also its tail and then the ratio $H_q$, by using a
superposition of two NBD's, but relaxing the constraint on the average 
multiplicities. The only constraint we impose is 
that the parameters $k$ of the two NBD's 
be the same, while we allow a variation of the difference between  
the average multiplicities. For the sake of simplicity and in order to be more
independent from any theoretical prejudice, we do not include any shift in the
MD for $b\bar b$ events. 

Formally, we perform then a fit 
with the following 3-parameter MD (plus the correction for the even-odd effect 
in eq.~\eref{pnfps}): 
\be
P_n(\bar n_l, \bar n_b, k) = \alpha 
P_n^{\mathrm{NB}}(\bar n_b, k) + (1 - \alpha ) P_n^{\mathrm{NB}}(\bar n_l, k)
\label{3par} 
\ee

The parameters of the fits and the corresponding $\chi^2$/NDF  
are given in Table~\ref{fits} (third group)
for different  values of the resolution parameter $\ymin$. 
A really accurate description of experimental data is achieved. 
Notice that the best-fit value for the difference between the average
multiplicities in the two samples, $\delta_{bl}$, 
also given in Table~\ref{fits}, is quite large. This difference
grows with increasing $\ymin$, i.e., with increasing contamination
of 3-jet events.

\begin{figure}
\begin{center}
\mbox{\begin{turn}{90}%
\epsfig{file=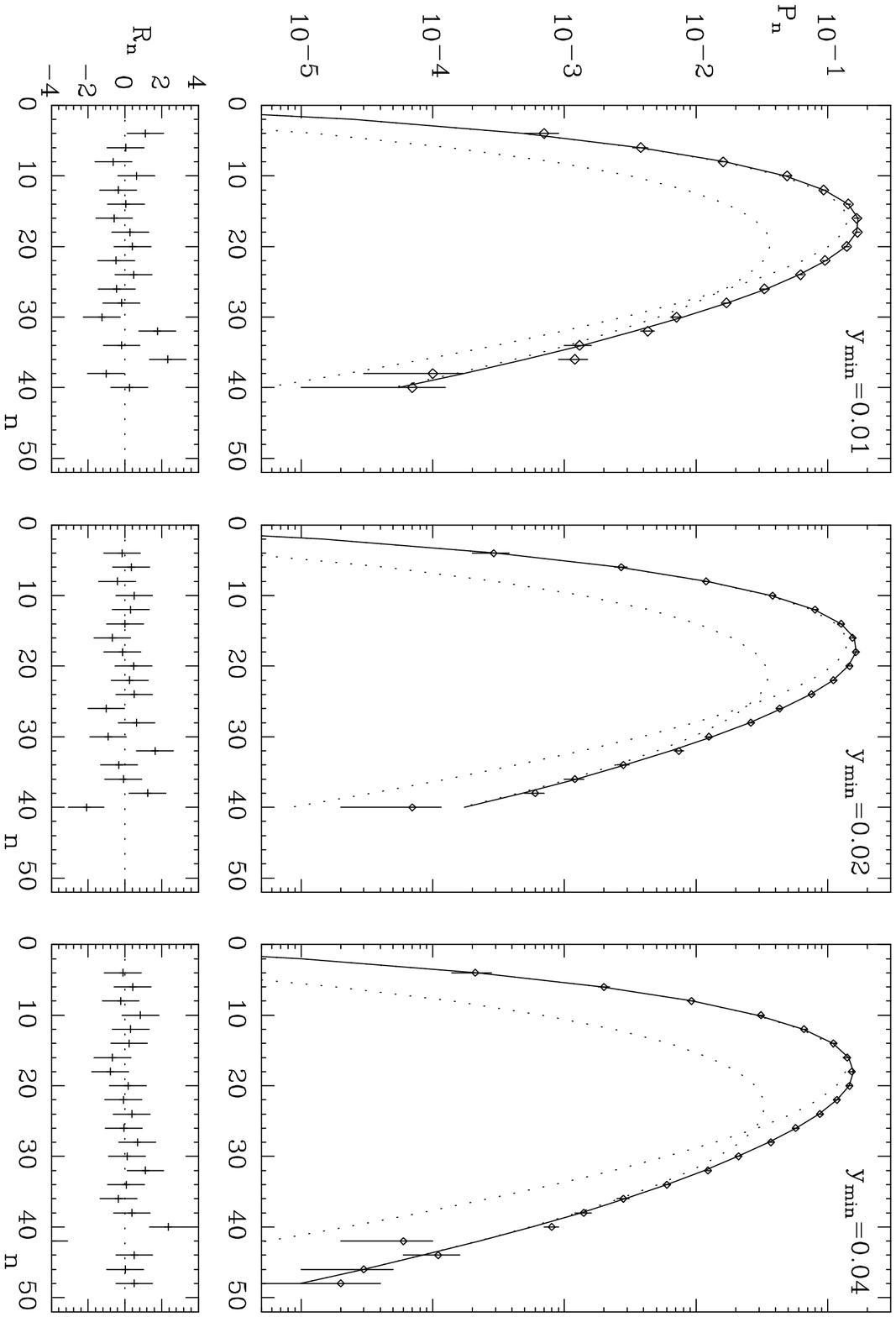,bbllx=0.cm,bblly=2.cm,bburx=22.cm,bbury=26.cm,width=15cm}
\end{turn}}
\end{center}
\caption[figure:fit3par]{
Charged particles' MD for two-jet events in full phase space, $P_n$, at the 
$Z^0$ peak  from
DELPHI\protect\cite{delphi:3} with different values of $\ymin$ are 
compared with a fit with the sum of 2 NBD's with the same parameter $k$
as in eq.~(\ref{3par}). The even-odd effect  has 
been taken into account (see eq.~\protect\eref{pnfps}). 
The lower part of the figure shows the residuals, $R_n$, i.e., 
the difference between data and theoretical predictions, 
expressed in units of standard deviations.} 
\label{fit3par}
\end{figure}

Figure~\eref{fit3par} compares the predictions of  eq.~\eref{3par} 
with the experimental MD's  for two-jet events at
different values of the resolution parameter $\ymin$. 
The residuals  are also shown  in units of standard deviations. 
 One concludes that 
the proposed parametrization can reproduce  the experimental data on MD's 
very well; no structure is visible in the residuals. 

As already discussed, the ratio $H_q$ gives a more stringent test of
theoretical parametrizations; it is then interesting to study the predictions
of eq.~\eref{3par} for this ratio. 
In this case, one can  obtain a closed expression for the factorial moments
 in terms of the parameters $\delta_{bl}$, $\bar n_l$ and $k$. 
Let us notice indeed that, since the two components are given by a NBD with the
same $k$, they have the same normalized factorial moments, which for a NBD are
given by: 
\be 
F_q^{(l)} = F_q^{(b)} = \prod_{i=1}^{q-1} \bigl( 1 + \frac{i}{k} \bigr)
\ee  
 From eq.~\eref{3par}, one obtains a similar relation for the generating
function: 
\be
G(z) = \alpha G^{(b)}(z) + (1-\alpha) G^{(l)}(z)
 \ee
 By differentiating 
 the previous equation, one then gets the following expression for 
 the unnormalized factorial moments, $\tilde F_q$: 
 \be
 \bar n = \tilde F_1 = \bar n_l + \alpha \delta_{bl}
 \ee
\ba
\tilde F_q &=& \alpha \tilde F_q^{(b)} + (1-\alpha) \tilde F_q^{(l)} = \\ 
&=& \biggl[ \alpha (\bar n_l + \delta_{bl})^q + (1-\alpha) 
{\bar n_l}^q \biggr]  \,\,
\prod_{i=1}^{q-1} \left( 1 + \frac{i}{k} \right) \nonumber 
\ea

\begin{figure}
\begin{center}
\mbox{\epsfig{file=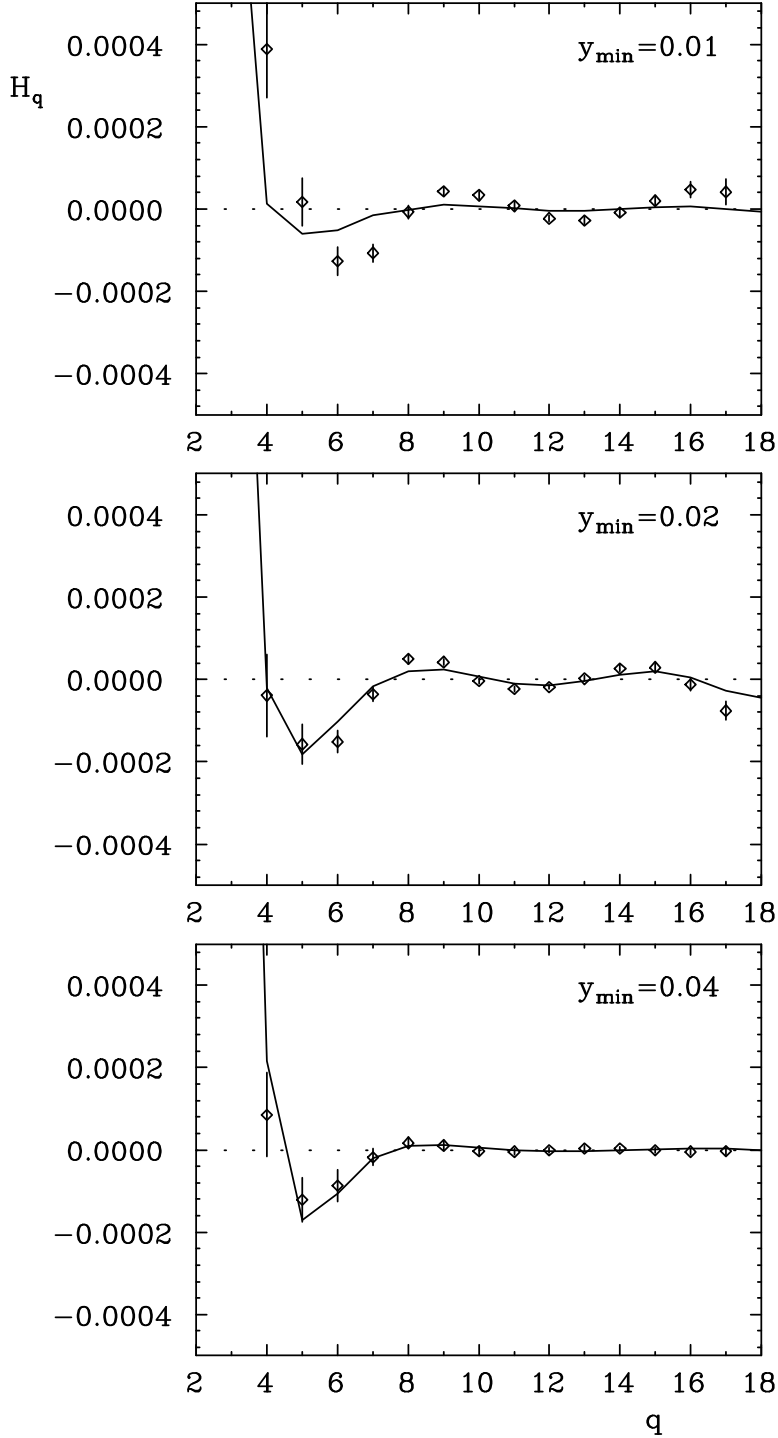,bbllx=4.cm,bblly=3.cm,bburx=16cm,bbury=27.cm,height=20cm}}          
\end{center} 
\caption[figure:hq2nbd]{The ratio of factorial cumulant 
over factorial moments, $H_q$, as a function of $q$; 
experimental data (diamonds) for 2-jet events 
with  different values of $\ymin$  
are compared  with equation~\protect\eref{3par} (solid lines). 
The parameters used are shown in Table~\protect\ref{fits}. 
The even-odd and the truncation effects have 
been taken into account (see eq.~\protect\eref{pntrunc}).} 
\label{hq2nb}
\end{figure}

Predictions of the ratio $H_q$ as a function of the order $q$ are obtained by
inserting eq.~\eref{3par} into eq.~\eref{pntrunc}. These predictions 
with parameters fitted to reproduce the MD as given in Table~\ref{fits} 
(third group) are compared in  Figure~\ref{hq2nb} 
with the $H_q$'s extracted from experimental data on MD's 
for 2-jet events\cite{delphi:3}  at different values of
$\ymin$  according to the procedure described in \cite{hq2}. 
The new parametrization gives an accurate description of the shape
of MD's and is shown to describe well the ratio $H_q$ too. 
Small deviations are still present for $\ymin=0.01$, where 
 2-jet events are  more collimated. They might be due to more subtle 
 not yet understood effects.  This consideration notwithstanding, 
 the overall description  of 2-jet MD's and $H_q$'s appears quite satisfactory. 
This result gives therefore  further support to 
the parametrization of the MD in quark-jets with fixed flavour 
in terms of a single NBD. 
It is also remarkable that  the average number of particles
only depends on flavour quantum numbers, whereas the NBD parameter $k$ is
flavour-independent. 

As a further check, we also investigated the MD of 2-jet
events with fixed flavour in the 
 Monte Carlo program \jetset\ 7.4 PS\cite{jetset}. 
For each flavour, we generated a sample of 60000 events and of 60000 2-jet
events (selected using the JADE algorithm with $\ymin=0.02$)
 by using the OPAL tuning\cite{opaltuning} and we fitted the MD's 
in full phase space with a
single NBD, eventually with a finite shift. 
In the all-events sample, the $\chi^2$/NDF is really bad, thus indicating that
a single NBD cannot describe the MD of events with fixed flavour. 
By requiring the 2-jet selection, the description improves strongly; 
the MD for light quarks are indeed well reproduced by a single NBD, 
while  $b\bar b$ events are better described by a shifted NBD.

\section{Conclusions} 

It has been shown that a single NBD cannot reproduce the observed behavior 
of the ratio $H_q$ for events with a fixed number of jets in full phase in
$e^+e^-$ annihilation at the $Z^0$ peak. 
A simple  phenomenological parametrization of the MD in terms 
of the weighted superposition of two  NBD's 
has been shown to describe
simultaneously both the MD's  and the ratio $H_q$. 
The weight of the first component was
taken equal to the fraction of  $b \bar b$ events, i.e., 
the two components were identified with the $b$- and 
the light flavours contribution respectively. 
The simple NBD parametrization 
is thus reestablished at the level of 2-jet events with fixed quark flavour
composition. 
It is interesting to note that 
this result is consistent with the results obtained in the context 
of a thermodynamical model of hadronization\cite{bgl}. 

It is remarkable that the two NBD's 
associated to $b$- and light flavours contributions 
have the same parameter $k$; 
since $k^{-1}$ is the second order normalized factorial cumulant, 
i.e., it is directly related to two-particle correlations, one concludes 
that two-particle correlations are flavour-independent in this approach. 
In addition, since both MD's are well described by a NBD, higher order 
correlations show a hierarchical structure\cite{LPA}, which is 
also flavour independent. 
This result can also be interpreted 
in the framework of clan structure analysis\cite{AGLVH:1}, where 
$k^{-1}$ gets the meaning of an aggregation coefficient, being 
 the ratio of the probability to have two particles in the
same clan over the probability  to have the two particles in two
separate clans: in this language, one concludes that 
the aggregation among particles produced into clans 
in $b\bar b$ and light flavoured events turns out to be  the same. 
The flavour quantum numbers affect then 
 the average multiplicity of the corresponding jets only, but 
 not the structure of particle correlations. 
 It would be interesting to see, when appropriate samples of events will 
be available, whether this  property established in
full phase space continues to be valid in restricted regions of it.

\section{Acknowledgements} 

Useful discussions with  W. Ochs  are gratefully acknowledged.

\end{document}